\documentstyle [12pt]{article}

\input [gwa.tex.pap]my_mac
\def\half{\frac{1}{2}}

\def\TC{T_{\scriptscriptstyle C} }

\def\Tone{T_{\scriptscriptstyle 1} }
\def\T2{T_{\scriptscriptstyle 2} }
\def\Tend{T_{\scriptscriptstyle \Omega} }

\def\Fc{F_{\scriptscriptstyle c} }

\def\leff{\lambda_{\scriptscriptstyle T} }

\def\vevphi{\langle \phi \rangle }
\def\phio{ \phi_{\scriptscriptstyle 0 } }
\def\phiplus{ \phi_{ + } }
\font\newsymb=msxm10   
                       \def\gapprox{\mathop{\hbox{\newsymb \&}}}
\begin{document}
\begin{titlepage}
\begin{center}
June   1992      \hfill     OHSTPY-HEP-T-92-009 \\

\vskip 1cm
{\Large
 {\bf Remarks on the Electroweak Phase Transition }}
\footnote{
To appear in {\it Proceedings, Yale Workshop on Baryon Number Violation
at the Electroweak Scale, March 1992} (World Scientific).  Work supported
in part by the U.S. Dept of Energy.}\\
\vskip 1.0 cm

       {
      {\bf Greg W. Anderson }\\
	\vskip .5 cm
       Physics Department, Ohio State University, 174 W. 18th Street\\
        Columbus, Ohio 43210\\  }

\vskip 2cm
\end{center}
\begin{abstract}
In the standard scenario, the electroweak phase transition
is a first order phase transition
which completes by the nucleation of critical bubbles.
Recently, there has been speculation that the standard picture of
the electroweak phase transition
is incorrect.   Instead, it has been proposed that throughout the
phase transition appreciable amounts of both broken and unbroken
phases of $SU(2)$ coexist in equilibrium.  I argue that this can not
be the case. General principles insure that
the universe will remain in  a homogenous state of unbroken $SU(2)$
until the onset of critical bubble production.

In addition, an analytic treatment
of the one Higgs doublet, electroweak phase
transition in the  standard model and  minimal extensions is reviewed.
Results from the thin wall approximation are compared to results obtained
using
the Lindes' action.   Perhaps the most important quantitative result we can
get from
an analysis  of the phase transition is determination of $\vevphi$ when
the phase transition completes.    For Higgs boson masses
above the current experimental limit, the thin wall approximation
determines the value of $\vevphi$ at the end of the phase transition
to an accuracy of better than three percent.

\end{abstract}
\end{titlepage}

\section{Introduction}
\indent

Spurred by the interest in electroweak baryogenesis,
a great deal of effort has been undertaken to describe and
quantify many salient aspects of the
electroweak phase transition (EWPT) in the minimal standard model
and its extensions.  It has become standard lore that the EWPT
is a first order phase transition which proceeds by the nucleation
and the subsequent growth of critical bubbles\cite{Dine,AH,Dinelinde}.
Although the phase transition is weakly first order in the
minimal standard model, the strength of the phase  transition is model
dependent.  Indeed, the EWPT is more strongly first order in very simple
extensions of the minimal standard model\cite{AH}.
The strength of the phase transition has been calculated analytically using
the thin wall approximation.
Section three of this note demonstrates that
the thin wall approximation gives highly reliable
values of $\vevphi$ at the end of the phase transition.

Recently however,  there has been speculation that the electroweak
phase transition is actually not a first order phase
transition after all\cite{KG}.
Instead, Gleiser and Kolb have suggested that  during the phase
transition the two phases of broken and unbroken $SU(2)\times U(1)$  coexist
simultaneously with equilibrium between the two phases being established and
maintained by sub-critical bubbles.  Similar arguments have been
advanced by Tetradis\cite{Nick}.
In particular, Gleiser and Kolb and Tetradis argue
that at the critical temperature $\TC$
the universe is filled with equal parts of the broken and unbroken
phases. $\TC$ is defined as the temperature of the universe when
the free energy
density of the system, plotted as a function of $\vevphi$ has two
degenerate minima.  The basic contention of these authors is that
as long as the expansion rate is slow compared to
the rates of thermal processes (Gleiser and Kolb consider
processes mediated by sub-critical bubbles), the universe will be driven into
a state equally populated by both phases.   If true, this argument would
have important ramifications for scenarios of baryogenesis which
invoke first order phase transitions.  In addition, one might wonder if
the standard picture of the universe trapped in a homogeneous
state when the temperature reaches $\TC$ is an assumption which has should
be checked case by case or if there are general dynamical and statistical
effects which guarantee this.
Below I will argue that the basic the picture advanced by Gleiser and Kolb is
in contradiction with the second law of thermodynamics.
Similar remarks would apply to the analysis of Tetradis and
previous studies of subcritical bubbles\cite{KGW}.
Instead, very general properties of statistical
mechanics guarantee that the equilibrium state at temperature $\TC$
is a homogeneous state.

\section{Collapsing Sub-Critical Bubbles}
\indent

Gleiser and Kolb and Tetradis argue that at the critical temperature
$\TC$, the universe is filled with equal parts of the broken and
unbroken phases.
The assertion that both wells are equally populated would be true
for an ensemble of particles interacting with an external potential.
However, metastability in one-dimensional
mechanics is very different from  metastability in a four-dimensional
field theory.   The disparate nature of these two cases is qualitative
as well as quantitative.  It is instructive to contrast these two
cases to see where some  types of intuition
can lead us astray.  Let's compare the symmetric double well  in
field theory and in one dimensional mechanics.\\

\vskip 5.5 cm
\begin{description}
{\small
\begin{center}
\item[ Figure 1:]  The symmetric double well in a mechanical
example  and in field theory.
\end{center}
}
\end{description}
First  consider the one dimensional mechanical example at fixed temperature.
For an ensemble of particles interacting with the
external potential given in figure 1a, the thermal equilibrium state
of the system is one where both wells are equally populated with particles.
This situation is in sharp contrast to the case in quantum field theory.
Under conditions present at the EWPT, the thermodynamic requirement
that the total entropy of the universe can  only increase is equivalent
to demanding that the free energy only decreases.  Thus,
the equilibrium state of the system minimizes the free energy.
Recall that the free energy of the system is:
\begin{equation}
F = \int d^3 x  \,\half (\nabla \phi)^2  + V(\phi,T).
\end{equation}
For convenience we can normalize the free energy density so that
$V(\phiplus) = V(\phio) = 0$.  A universe held at a temperature $\TC$
and left to equilibrate will end up either
in the homogeneous ground state
$\vevphi = \phi_{+}$  or $\vevphi = \phio = 0$.   This is
because both the gradient
term and $V(\phi,T)$ are positive and nonvanishing inside any boundary
separating the two
phases.   So a universe filled with domains
of both phases must have a larger free energy than either homogeneous
state.  According to the second law of thermodynamics,
while an individual fluctuation can
occasionally increase the systems free energy, the cumulative
statistical effect of these
fluctuations  must decreases the
free energy. Moreover, because the universe is cooling, there is no question
which state the universe occupies when the temperature
reaches $\TC$.
The state of lowest free energy at temperatures above $\TC$ is
$\vevphi = \phio$, and when the temperature reaches $\TC$ the newly degenerate
homogeneous vacua at $\vevphi = \phi_{+}$ is
separated from the state $\vevphi = \phio$ by an infinite
barrier.\footnote{
At temperatures above $\TC$, the metastable state
which first appears at $\vevphi = \phi_{+}$ is separated from
the homogeneous state $\vevphi = \phio$ by an infinite barrier.
Any finite {\em finite} region of space containing the new phase is not a
meta-stable state since it can be continuously deformed to the
ground state without surmounting an energy barrier.  At temperatures equal to
and
above $\TC$ finite regions of $\vevphi = \phiplus$  are completely unstable.
Only when the
temperature drops below $\TC$, the can system can be continuously deformed from
the state $\vevphi = \phio$ to the new equilibrium state
$\vevphi = \phiplus$ by crossing a finite
barrier (see figure 2).
The height of this free energy barrier is the critical bubble
free energy.}
So when the temperature drops to $\TC$
the universe finds itself in the homogeneous vacuum $\vevphi = \phio$.
This is in direct conflict with the analyses of Gleiser and Kolb
and Tetradis.

Although it is clear from these general grounds that the universe
is filled with the homogeneous state $\vevphi = 0$
when the temperature cools to $\TC$ ,
it is useful to  see the how this arises from the dynamical
equations governing the evolution of the scalar field.
This will allow us to quantify how large
fluctuations are about the equilibrium state.
Before discussing the statistical evolution of  the scalar field
it will be useful to
recall a few basic properties of nucleated bubbles.
Consider a bubble containing $\vevphi = \phiplus$  in a sea of vacuum
$\vevphi = 0$ (see figure 2).
By convention we will choose the state $\vevphi = 0$ to have
zero free energy.
\vspace*{ 5 cm}
\begin{description}
{\small
\begin{center}
\item[ Figure 2:]  The effective potential at temperatures near $\TC$
\end{center} }
\end{description}

Then the surplus free energy of a nucleated bubble is
\begin{equation}
F = \int d^{3}x
\left\{
\half\left( \vec{ \nabla } \phi \right)^{2}
+ V(\phi,T)  \right\}.
\end{equation}
The free energy of this bubble has two contributions:
a surface free energy $F_{S}$, coming mostly from the derivative terms
in Eq. 2.2,
and a volume term $F_{V}$, which arises  from the difference in
free energy density inside and outside the bubble.
These two contributions
scale like
\begin{eqnarray}
 F_{S}&\sim & \,\,\,2\pi R^{2}
\left( \frac{\delta \phi}{\delta R}\right)^{2} \delta R, \nonumber\\
F_{V}&\sim & \, \frac{4\pi}{3}\,R^{3} \,\bar{V}(\phiplus) ,
\end{eqnarray}
where $R$ is the radius of the bubble, $\delta R$ is the thickness of
the bubble wall, $\delta \phi \sim \phiplus $,
and $\bar{V}(\phiplus)$ is the average value of the potential inside
the bubble.
 For the bubbles we are interested in, it is energetically
favorable to make the gradient term as small as possible, so the bubbles
will be thick walled.
 For thick walled bubbles
$\delta R \sim R$, and the surface energy of the
bubble grows like $R$. In contrast, the volume term increases in
magnitude like $R^{3}$.
For temperatures below $\TC$ the volume term in equation Eq. 2.3 can
be negative (See figure 2).  At this temperature,
although the homogeneous state $\vevphi = \phiplus$
has a lower free energy, a thermal fluctuation
producing a bubble
of true vacuum which starts from a  radius of zero and expands in radius
to envelope the system, must have a free energy greater than
or equal to some critical value.  The radius of
the {\em critical} bubble
is found by  differentiating Eq. 2.3,
 $R_{c}\sim \phiplus/\sqrt{-2\bar{V}(\phiplus)}$.
Sub-critical bubbles, those bubbles with radii smaller than this
critical size,  will collapse
under their surface tension. The free energy of a
critical bubble is:
\begin{equation}
F_{c} \sim
\frac{ \phiplus^{3}}{\sqrt{-\bar{V}(\phiplus)}}.\\
\end{equation}
Notice as the temperature approaches the
critical temperature from below $\bar{V}\rightarrow 0$, and both the
radius and free energy of the critical bubble become infinite.  So
at $T\geq \TC$, all bubbles are subcritical.

   Aided by this qualitative understanding of nucleated bubbles
we can examine the dynamical equations describing the abundance of
regions containing the $SU(2)$ broken phase.
In a hot universe in the vacuum state $\vevphi =\phio = 0$,
thermal fluctuations will produce bubbles inside of which
$\vevphi$ is nonvanishing. These
thermal fluctuations produce critical bubbles at a rate per unit volume:
\begin{equation}
\Gamma /V \simeq T^4 e^{-\beta \Fc}.
\end{equation}
The exponential suppression $\exp(-\beta \Fc)$ is the
usual Boltzmann suppression
for producing configurations close to the
critical bubble.
This suppression arises because the system must cross
a barrier in order to produce bubbles large enough to grow,
while the prefactor gives the rate of typical
processes which are not Boltzmann suppressed.
Let $f$ denote the fraction of space filled regions of $\vevphi \sim
\phi_{+}$.  The master equation for the evolution of $f$ is:
\begin{equation}
\frac{df}{dt} = \left(1 - f\right)\Gamma(\phio\rightarrow\phiplus)
 - f \Gamma(\phiplus \rightarrow \phio)
\end{equation}
 Although the rate given in Eq. 2.5 has strictly only been motivated for
critical bubble production,
it is not unreasonable to assume that its generalization gives a
good estimate of the rate other configurations are produced.
Any fluctuation producing a region of $\vevphi \sim \phiplus$,
will be Boltzmann suppressed because energy is required to form
 the domain boundaries. So the regions of $\vevphi \sim \phio$
with spatial extent $R$ will be converted to regions of
 $\vevphi \sim \phiplus$ at a rate:
 \begin{equation}
\Gamma(\phiplus \rightarrow \phio) \simeq T(RT)^3  e^{-\beta F},
\end{equation}
where $F$ is the free energy of a subcritical bubble of radius $R$.
{}From Eq. 2.3, $F \gapprox 2\pi \phiplus^2 R$.
Fluctuations which create energetically
disfavored structures can also remove them.
Even in a universe filled equally with domains of both phases,
there will be fluctuations which decrease the abundance
of domain walls and take the universe toward
a homogeneous state.  Fluctuations of this sort, which decrease
the volume occupied by domain boundaries, do not cost energy
so their rates are not Boltzmann suppressed.
If anything they should be enhanced relative to rates which leave
$\vevphi$ unchanged.
Regions of $\vevphi \sim \phiplus$ with spatial extent $R$
are depleted by both fluctuations, and the dynamical collapse resulting
from the region's surface tension:
\begin{equation}
\Gamma(\phio \rightarrow \phiplus) \gapprox T(RT)^3 + 1/\tau .
\end{equation}
A simple estimate of the collapse time gives $\tau \sim R$.
Although the self induced collapse is typically  faster than
the rate of bubble production, fluctuations are even more
effective at removing regions of the unstable phase.
In steady state, detailed balance requires that the fraction of space
containing bubbles of broken phase is exponentially suppressed.
{}From Eq. 2.6 - 2.8,
\begin{equation}
f  \leq  \frac{ e^{-F/T} }{1 + e^{-F/T} }.
\end{equation}
where $F$ is the free energy of a sub-critical bubble including the
bubble walls.
Since we are interested in bubbles which change the value of the
scalar field condensate we can set a lower limit on the
magnitude of a bubbles free energy.
In order to treat the scalar field condensate classically, a bubble
of scalar field must contain many quanta\cite{AHH}.
Since the wavelength of a typical quantum comprising a bubble is
order $R$, with $F \sim n_{q}/R$ and $n_{q}>>1$ we must have
$(F/T) (RT)>>1$, where $n_{q}$ is the number of quanta.  Using Eq. 2.3,
\begin{equation}
F/T  \gapprox 2\pi \left(\frac{\phiplus}{T} \right)^2 (RT) >> 1/RT.
\end{equation}
In the standard model, the ratio $F/T$ satisfying Eq. 2.10 is
not small.
Thus, until the onset of critical bubble nucleation,
 the universe finds itself in a homogeneous state with
an exponentially suppressed number of ephemeral regions
containing  the $SU(2)$ broken phase.

Although Eq. 2.9 demonstrates that before the phase
transition occurs, the unbroken phase
is always favored, one might wonder if there are models where Eq. 2.9
allows for a departure from the standard formalism of false vacuum
decay.  In models where $\phiplus/T <<1$ it is possible to have
$FR>>1$ without $F/T<<1$.  Indeed, we know  in the limit $\phiplus
\rightarrow 0$ we must have $f \rightarrow \half$.  However, even
in the extreme case $\phiplus/T <<1$ the standard formalism
of first order phase transitions should remain valid.  The new ground state
$\vevphi = \phiplus$ will not dominate until fluctuations can
produce regions large enough to grow, and this will not happen until
the temperature drops below $\TC$.  Whether such a region is produced
all at once or by the coalescence of smaller regions, the rate for
producing the saddle point solution is  given by Eq. 2.5.
\footnote{ When calculating the thermodynamic probability of producing a
critical bubble by a saddle point evaluation of the partition
function no choice is made to include some histories at the expense of
 others. }
The first order phase transition will occur at a temperature where
fluctuations produce regions of the new phase  large enough
to grow at a rate which exceeds the expansion rate of the universe.

\section{The Thin Wall Approximation and the EWPT}

\indent

A completely analytic treatment of the EWPT has been given using the
thin wall approximation\cite{AH}.
An analytic treatment of the
phase transition  makes apparent simple ways in which
the standard model can be augmented
to prevent washout of the baryon asymmetry
after completion of the  EWPT.  Recently, an analytic fit to
the free energy of a critical bubble was devised\cite{Dinelinde}.
Using the Lindes' action we can quantify the utility of the thin
wall approximation used in reference 2, and
 analytic results  can be extended
to include quantities
and domains where the thin wall approximation is less reliable
\cite{Dinelinde}.

At temperatures near the electroweak phase
transition, the effective potential for the standard model
can be reliably written as a polynomial in $\phi$\cite{AH,Sher}:
\begin{equation}
V(\phi,T)  =D(T^2-\T2^2)\phi^2 - ET\phi^3 + \frac{1}{4}\leff \phi^4.
\end{equation}
The relationship between the constants $D,E,\T2,\leff$ at one loop, and
standard model
parameters can be found in reference 2.   Higher order effects decrease
the value of $E$ by a factor of $2/3$ relative to the one loop result
making the phase transition more weakly first order\cite{Dinelinde}.
$\T2$, the temperature at which the
origin becomes unstable,
is on the order of the Higgs Boson mass.
At temperatures
far above $\T2$, the only minimum of the potential is
$\vevphi=0$.
As the early universe cools down from this
high temperature, a second local minimum of the potential
appears, and later becomes degenerate with the origin.  This
occurs  when
the temperature drops to $\Tone$, which is related to $\T2$ by:
\begin{equation}
\Tone^2 = \frac{1}{1-\frac{E^2}{\leff D} }\,\T2^2.
\end{equation}
The $SU(2)$ breaking, local minimum $\phiplus$ evolves as
\begin{equation}
\phi_{+}=\frac{3ET}{2\leff} + \frac{1}{2\leff}
\sqrt{ 9E^2T^2 - 8 \leff D\left(T^2-\T2^2\right) }.
\end{equation}

A tenable scenario for baryogenesis
at the electroweak phase transition must ensure that
the rate of baryon number violation is suppressed at the end of the phase
transition.   This condition can be translated into a limit on the mass
of the Higgs Boson.  An accurate bound on the effective Higgs Boson
mass requires a precise determination
of $\vevphi$ at the end of the phase transition.
Indeed, $\phi_{+}/T$ increases by a factor $3/2$ as the temperature
drops from $\Tone$ to $\T2$, so the Higgs mass limit can in principle change
by more than $20$ percent depending on when the phase transition
completes.

 In order to determine when the phase transition completes
we need the free energy of a critical bubble.
Defining $\epsilon = (\Tone -T)/(\Tone-\T2)$,
rescaling both the spatial coordinates and the scalar field,
with the potential given by Eq. 3.1, for $\Tone +\T2 >> \Tone - \T2$
the free energy of a critical bubble  can be written:\cite{AH}
\begin{equation}
\Fc(T)/T  = \left(\frac{64\pi}{81}\right)
\frac{E}{(2\leff)^{3/2}}  f(\epsilon),
\end{equation}
where $f$ is function which depends only on $\epsilon$.  In
the thin wall approximation $f = \epsilon^{-2}$.  In reference 3
an analytic function was obtain which reproduces the numerical
value of the free energy to within a few percent.
Rescaling the Lindes' function and writing it in terms of epsilon
,\begin{eqnarray}
 f(\epsilon) &=&15. (1-\epsilon)^{3/2}\left[1+\frac{1-\epsilon}{4}
     \left(1 + 2.4/\epsilon + .26/\epsilon^2 \right) \right] \nonumber \\
          &= &
         (1 + 6.7\epsilon -2.0\epsilon^2 - 16.\epsilon^3 + ...)/\epsilon^2,
\end{eqnarray}
which reproduces the result of the thin wall approximation as $\epsilon
\rightarrow 0$.  With the bubble nucleation rate given by Eq. 2.5,
the phase transition completes when
\begin{equation}
\Fc(\Tend)/\Tend \simeq 100.
\end{equation}
Typically,
the phase transition completes soon after the universe
cools to temperature $\Tone$.  The values
of $\epsilon_{\Omega}$ and $\phi_{+}(\Tend)/\Tend$
are shown as a function of the
Higgs boson mass with the Lindes' action and with the thin wall approximation
in figures 3 and 4.  The result is plotted for a top quark mass of
$\mt = 120.$  Larger values of the top quark mass increase the reliability
of the thin wall approximation.
\newpage
\vspace*{ 8cm}
\begin{description}
{\small
\item [Figure 3:]  Epsilon at the end of the phase transition
verses the Higgs boson mass for  for $\mt = 120$.  The dashed curve
is the thin wall approximation, while the solid curve is the
result using the Lindes' action.
}
\end{description}

\vspace*{8.5cm}
\begin{description}
{\small
\item[ Figure 4:]  $\phiplus$ at the end of the phase transition
verses the Higgs boson mass for $\mt = 120$.   The dashed curve
is the thin wall approximation, while the solid curve is the
result using the Lindes' action.  The upper and lower dotted curves
correspond to the values of $\phiplus/T$ at temperatures $\T2$ and $\Tone$
respectively.}
\end{description}

In simple extension of the standard model, virtual effects of
additional particles can noticeably change
 the relationship between $\leff$ and
the Higgs boson  mass\cite{AH}.
In some cases, the effect of these
interactions makes the phase transition proceed  as it would if
the Higgs Boson mass was significantly smaller\cite{AH}.
 For this reason
the graphs of figures 4 and 5 have been extended below the current
experimental limit on the Higgs Boson mass.  Note that for $\mH =
30 \,(60)$ GeV, the thin wall approximation determines $\phiplus/T$
to an accuracy of $6.5 \,\,(2.5)$ percent, while the error in $\epsilon$
is 38 (27) percent.

Why should we be interested in extensions of the standard model which
produce these virtual effects?
Even if the standard model is augmented with enough additional $CP$
violation to
make baryogenesis during the EWPT tenable, we are still
left with the problem of how to avoid depletion of the $B+L$ asymmetry
just after $\Tend$ as soon as thermal equilibrium is re-established.
Several authors \cite{Dine,shap}
 have argued that anomalous baryon
number violation  will washout  any baryon asymmetry for Higgs masses larger
than some critical value, $m_{H_c}$.
Although there are uncertainties in the calculation of $m_{H_c}$,
it is in the vicinity of 35 GeV, well below the experimental lower bound.
Roughly, this corresponds to
\begin{equation}
\frac{ \phiplus( \Tend )}{\Tend} \simeq
2 \left({E\over \lambda_{T}}\right)_c \simeq 1
\end{equation}
However, particles can be added to the standard model
so that $E/\leff > \half$
for any desired value of $m_H$. This can be accomplished  by
adding bosons with small $SU(2) \times U(1)$ preserving masses
to increase $E$, or by adding bosons so that $\leff/\mH^2$ is
decreased\cite{AH}.
So  successful baryogenesis at the one Higgs doublet EWPT can
occur provided the Higgs doublet is given two new interactions: one
to violate CP and the other to enhance $\vevphi$ at
the end of the phase transition.
A demonstration
of the utility of the thin wall approximation is the accuracy with
which $\vevphi$ at the end of the phase transition is obtained.

\end{document}